\documentclass[prl,onecolumn,preprintnumbers,amsmath,amssymb,11pt]{revtex4}
\usepackage[utf8]{inputenc}
\usepackage{epsf}
\usepackage{graphicx}
\usepackage{latexsym}
\usepackage{amsmath}
\usepackage{amssymb}
\usepackage{amsfonts}
\usepackage{color}

\begin{document}

\title{\Large{Field-Emission Resonances in Thin Metallic Films: Nonexponential Decay of the Tunneling Current as a Function of the Sample-to-Tip Distance}}


\author{Alexey Yu. Aladyshkin$^{(a,b,c,*)}$ and Koen Schouteden$^{(d)}$}

\medskip
\affiliation{$^{(a)}$Institute for Physics of Microstructures RAS, GSP-105, Nizhny Novgorod, 603950 Russia \\
$^{(b)}$Lobachevsky State University of Nizhny Novgorod, Gagarin Av. 23, Nizhny Novgorod, 603022 Russia \\
$^{(c)}$National Research University Higher School of Economics (HSE University), Myasnitskaya str. 20, 101000 Moscow, Russia\\
$^{(d)}$Semiconductor Physics Section, Katholieke Universiteit Leuven, Celestijnenlaan 200D, B-3001 Leuven, Belgium}

\date{\today}

\maketitle




\section{Abstract}

Field-emission resonances (FERs) for two-dimensional Pb(111) islands grown on \mbox{Si(111)7$\times$7} surfaces were studied by  low-temperature scanning tunneling microscopy and spectroscopy (STM/STS) in a broad range of tunneling conditions with both active and disabled feedback loop. These FERs exist at quantized sample-to-tip distances $Z^{\,}_n$ above the sample surface, where $n$ is the serial number of the FER state. By recording the trajectory of the STM tip during ramping of the bias voltage $U$ (while keeping the tunneling current $I$ fixed), we obtain the set of the $Z^{\,}_n$ values corresponding to local maxima in the derived $dZ/dU(U)$ spectra. This way, the continuous evolution of $Z^{\,}_n$ as a function of $U$ for all FERs was investigated by STS experiments with active feedback loop for different $I$. Complementing these measurements by current--distance spectroscopy at a fixed $U$, we could construct a 4-dimensional $I-U-Z-dZ/dU$ diagram, that allows us to investigate the geometric localization of the FERs above the surface. We demonstrate that (i) the difference $\delta Z^{\,}_n=Z^{\,}_{n+1}-Z^{\,}_n$ between neighboring FER lines in the $Z-U$ diagram is independent of $n$ for higher resonances, (ii) the $\delta Z^{\,}_{n}$ value decreases as $U$ increases; (iii) the quantized FER states lead to the \emph{periodic} variations of $\ln I$ as a function of $Z$ with periodicity $\delta Z$; (iv) the periodic variations in the $\ln I - Z$ spectra allows to estimate the absolute height of the tip above the sample surface. Our findings contribute to a deeper understanding on how the FER states affect various types of tunneling spectroscopy experiments and how they lead to a non-exponential decay of the tunneling current as a function of $Z$ at high bias voltages in the regime of quantized electron emission.

\medskip
$^*$ Corresponding author, e-mail address: aladyshkin@ipmras.ru

\section{Introduction}

An appearance of surface electronic states localized near flat surfaces of crystals was predicted theoretically by Tamm \cite{Tamm-JETP-32} in the tight-binding model and later by Goodwin \cite{Goodwin-Proc-35} and Shockley \cite{Shockley-JETP-39} within the nearly-free-electron approximation. It was demonstrated that the discontinuity of the potential energy at the crystalline surface could generate stationary localized solutions of the Schr\"{o}dinger equation corresponding to exponentially decaying wave functions both in vacuum above the sample surface and  in the bulk below the sample surface (with atomic-scale oscillations). A review of various methods of theoretical analysis of surface electronic states is presented in Ref.~\cite{Davison-book}.

Rather than a discrete discontinuity, in reality the electric potential above the metallic crystal surface has a slowly decreasing component ($\sim 1/z$, where $z$ is the distance from the metal-vacuum interface) because of the electrostatic Coulomb interaction between the probe electron and its mirror (image) charge (see, \emph{e.g.} \cite{Book-Electrodynamics-1,Book-Electrodynamics-2}). The appearance of the Coulomb-like potential could lead to the localization of electrons above the surface of conducting sample provided that the energies of these states correspond to the forbidden band of bulk crystal. The effect of trapping of electrons above the flat surface is similar to the localization of electrons in hydrogen-like atoms, and therefore it leads to a Rydberg-like spectrum of excitations (see reviews \cite{McRae-RMP-79,Echenique-review-1,Echenique-review-2} and references therein). In literature these electronic states are known as image-potential states (IPSs), although from physical point of view these states can be considered as a generalization of the Tamm-Shockley surface states for the case of continuously varied electrical potential near the surface, what results in a shift of the oscillatory part of the wave function outside the sample (figures\,S1 and S2 in Supporting Information). IPSs in the absence of external electric field are usually studied by means of photoemission spectroscopy \cite{Garcia-PRL-85,Hofer-Science-97}.

\begin{figure*}[t!]
\centering{\includegraphics[width=14.5cm]{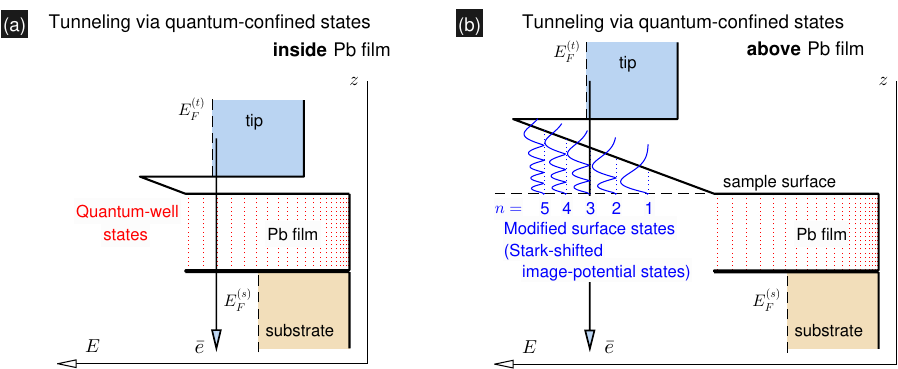}}
\caption{ (color online)
({\bf a, b}) Schematic presentation of the resonant processes of electronic coherent transport from the STM tip to the sample (the Pb(111) film on the Si(111) substrate) through quasi-stationary electronic states localized (i) inside the two-dimensional metallic film at low bias voltages (panel a) and (ii) above the metallic film at high bias voltages (panel b) for constant tunneling current. The effect of the image potential on the shape of the potential barrier is not shown. Yellow and blue rectangles show the filled electronic states for the substrate and the tip, correspondingly; $\varphi^{\,}_s$ and $\varphi^{\,}_t$ are electrical potentials of the substrate and the tip, correspondingly; $E^{(s)}_F=E^{\,}_F-|e|\varphi^{\,}_s$ and $E^{(t)}_F=E^{\,}_F-|e|\varphi^{\,}_t$ are the electrochemical potentials of the sample  and the tip, respectively;  $U=\varphi^{\,}_s-\varphi^{\,}_t>0$ is the bias voltage, $n$ is the FER index equal to the number of electronic half-waves above the sample surface. Adapted from Ref.~\cite{Aladyshkin-JPCC-21} (Aladyshkin et al. {\it J. Phys. Chem. C} {\bf 2021}, {\it 125}, 26814). Copyright 2021 American Chemical Society.}
\label{Fig01}
\end{figure*}

The effect of quasi-stationary electronic states localized above a surface of conducting samples on transport properties of quantum-confined system becomes more pronounced in the presence of external sources of electric field. This configuration is typical for scanning tunneling microscopy (STM) and spectroscopy (STS). Indeed, for a non-zero tunneling voltage (bias) $U$ between a STM tip and a sample the effective electric potential near the surface can be viewed as a superposition of the Coulomb potential of the image charge and the linearly increasing function of $z$. Obviously, the positions of all modified surface states in the presence of the external electric field are shifted upward to higher energies \cite{Binnig-PRL-85} similar to the Stark effect for a hydrogen-like atom \cite{Landau-III}. The wave functions corresponding to the modified surface states in linearly increasing electric potential are shown schematically in figure~\ref{Fig01}b. The influence of the quantized electronic states localized within potential barrier area on the energy dependence of the transmission coefficient for the trapezoidal tunneling barrier and on the current-voltage ($I-U$) dependence were theoretically considered by Gundlach \cite{Gundlach-SolStateElectr-66}. Sometimes the oscillations of the differential conductance $dI/dU$ as a function of $U$ conditioned by resonant tunneling via the modified surface states are referred to as the Gundlach oscillations or field-emission resonances (FERs). The FERs were experimentally studied for atomically flat surfaces
\cite{Becker-PRL-85,Bobrov-Nature-01,Wahl-PRL-03,Jung-PRL-95,Hanuschkin-PRB-07,Kubetzka-APL-07,Silkin-PRB-09,Jarvinen-JPCC-2015,Martinez-Blanco-PRB-15,Ge-PRB-20, Borca-2D-20};
and for various nanostructured objects on top of flat surfaces
\cite{Borisov-PRB-07,Rienks-PRB-05,Dougherty-PRL-06,Pivetta-PRB-05,Ploigt-PRB-07,Lin-PRL-07,Aladyshkin-JPCM-20,Konig-JPCC-09,Schouteden-PRL-09,Liu-Nanolett-21,Stuckenholz-JPCC-15,Schouteden-PRL-12,Yang-PRL-09,Becker-PRB-10,Schouteden-Nanotech-10,Craes-PRL-13,Sugawara-PRB-17,Park-JPCC-12}. In particular, the analysis of the spectra of the FERs allows to estimate the local work function \cite{Pivetta-PRB-05,Ploigt-PRB-07,Lin-PRL-07,Aladyshkin-JPCM-20,Konig-JPCC-09,Stuckenholz-JPCC-15,Liu-Nanolett-21,Kolesnychenko-RSI-99,Kolesnychenko-PhysB-00} and to probe the profile of surface potential \cite{Bono-SurfSci-87,Ruffieux-PRL-09}.

In this paper we study geometric localization of FERs for thin Pb(111) films on \mbox{Si(111)$7\times 7$} surfaces by means of low-temperature STS in the regime of constant tunneling current $I$ and variable distance $Z$ between the STM tip and the Pb surface. The main aim of the paper is to demonstrate that the FERs affect strongly the current--distance ($I-Z$) relationship, resulting in periodic variations in the logarithmic derivative of $I$ with respect to $Z$ (\emph{i.e.} $d(\ln I)/dZ\equiv I^{-1} dI/dZ$) as a function of $Z$ for a fixed tunneling voltage. This observation apparently differs from well-known exponential-like dependences of the tunneling current as a function of the width of the low-transmission potential barrier in the limit of low and high voltages. Indeed, the dominant contribution to $d (\ln I)/dZ$ for planar contacts at large distances is controlled by the mean work functions $\bar{W}=(W^{\,}_s+W^{\,}_t)/2$ of the sample and the tip \cite{Holm,Simmons-1,Simmons-2}
\begin{eqnarray}
\label{Eq01}
\frac{d(\ln I)}{dZ} \simeq - 2 \frac{\sqrt{2m^{\,}_0}}{\hbar}\,\bar{W}^{1/2} \quad \mbox{at} \quad |eU|\ll \bar{W},
\end{eqnarray}
or by the work function of the tip $W^{\,}_t$ \cite{Holm,Simmons-1,Simmons-2}
\begin{eqnarray}
\label{Eq02}
\frac{d(\ln I)}{dZ} \simeq - \frac{4}{3}\,\frac{\sqrt{2m^{\,}_0}}{\hbar}\,\frac{W^{3/2}_t}{|eU|} \quad \mbox{at} \quad |eU|\gg \bar{W}.
\end{eqnarray}
Here $m^{\,}_0$ and $e$ are mass and charge of electron in a vacuum spacer, $U$ is the mean bias voltage, $W^{\,}_s$ and $W^{\,}_t$ are work functions for the sample (collector of electrons) and the tip (emitter of electrons). Important to note that both estimates (\ref{Eq01}) and (\ref{Eq02}) are independent on $Z$, what should correspond to universal linear decrease of $\ln I$ as function of $Z$ regardless on the bias. However we find out that trivial exponential decay of tunneling current or, equivalently, linear decay of $\ln I$ as a function of $Z$ for large bias voltage is substantially modified in the regime of quantized electronic emission. We determine the set of quantized heights $Z^{\,}_n$ and energies $|e|U^{\,}_n$, corresponding to the local maxima of $\left(dZ/dU\right)^{\,}_{I={\rm const}}$, that are related to the FER states, where $n=1,2,\ldots$ is the serial number of the resonance. We argue that the period of the oscillations of the logarithmic derivative of the tunneling current $\delta Z^{\,}_n = Z^{\,}_{n+1}-Z^{\,}_{n}$ is independent on $n$ and monotonously decreases as $U$ increases. To the best of our knowledge, such bias-dependent periodic oscillations of $d(\ln I)/dZ$ as a function of $Z$ in STS experiments have not been yet discussed in literature.
Our experimental findings can be interpreted within the quasi-classical model for electron trapped in the triangular potential well. Based on our experiments and analysis, we can state that similar effects controlled by resonant tunneling through quasi-stationary states above the sample surface should exist for any conducting material provided that the STM tip has a proper shape.

\begin{figure*}[t!]
\centering{\includegraphics[width=8cm]{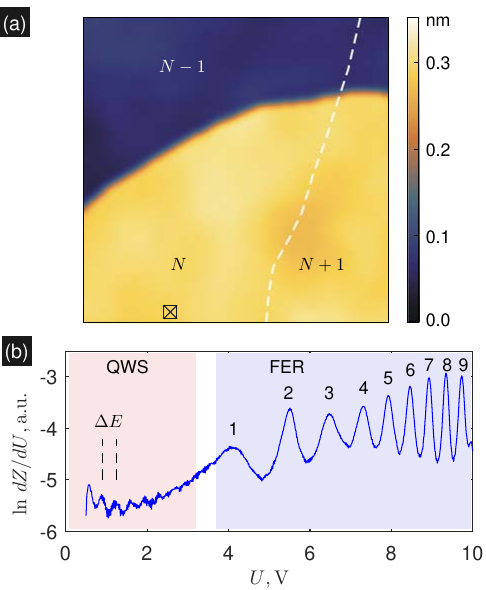}}
\caption{(color online) {\bf (a)} Aligned topography image $z(x,y)$ of the surface of the Pb(111) island (image size is $58\times 58$~nm$^2$, $U=0.5\,$V, $I=200\,$pA). The dashed line indicates the position of the hidden monatomic step in the Si(111)$7\times 7$ substrate (the method of visualizing of the hidden defects in Pb films was described in \cite{Putilov-JETPLett-19,Ustavshchikov-JETPLett-17,Aladyshkin-JPCC-21}). The local thickness $D$ of the island in the left lower corner (at the point $\boxtimes$) is about 10.9\,nm with respect to the Si(111)$7\times 7$ surface (i.e. $N=D/d^{\,}_{ML}\simeq 38$, $d^{\,}_{ML}=0.286$\,nm is the thickness of a Pb monolayer for the Pb(111) surface).\\
{\bf (b)} Typical dependence of $dZ/dU$ on the bias potential $U$, acquired for $I=1600\,$pA at the point $\boxtimes$.}
\label{Fig02}
\end{figure*}

\section{Methods}

Experimental investigations of electrophysical properties of quasi-two-dimensional Pb islands were carried out in an ultra-high vacuum (UHV) low-temperature scanning probe microscopy setup (Omicron Nanotechnology GmbH) operating at a base vacuum pressure \mbox{$2\cdot10^{-10}\,$mbar}. Si(111) crystals were first out-gassed at about 600$^{\circ}$C for several hours then cleaned \emph{in-situ} by direct-current annealing at about 1300$^{\circ}$C, resulting in formation of reconstructed Si(111)7$\times$7 surface. Thermal deposition of Pb (Alfa Aesar, purity of 99.99\%) from a Mo crucible was performed {\it in-situ} on Si(111)7$\times$7 surface at room temperature by means of an electron-beam evaporator (Focus GmbH, model EFM3) at \mbox{$6\cdot10^{-10}\,$mbar}. The orientation of the atomically flat terraces at the upper surface of the Pb islands corresponds to the (111) plane \cite{Altfeder-PRL-97,Su-PRL-01,Eom-PRL-06}. All STM/STS measurements were carried out at liquid nitrogen temperatures (from 78 to 81\,K) with electrochemically etched W tips cleaned {\it in-situ} by electron bombardment.
To optimize the shape of the STM tip apex after the cleaning, we touch the surface of clean Pb terraces by changing the distance between the tip and the sample surface from zero (actual position) to $-4\,$nm (well below surface). We can therefore assume that our W tips become covered by a Pb layer during our series of experiments. The lateral and vertical thermal drifts of the piezo-scanner after thermal stabilization are less than 0.5~nm/hour.

The topography of the Pb islands was studied by low-temperature STM by tracking the displacement of piezoscanner $z(x,y)$ during scanning along the sample surface with active feedback loop (constant tunneling current $I$) and constant electrical potential of the sample ($\varphi^{\,}_s=U$) with respect to a virtually grounded STM tip ($\varphi^{\,}_t=0$). In order to study the variations of the electronic properties of the sample at sweeping $U$ we use the following measurement protocol \cite{Schouteden-PRL-09,Schouteden-PRL-12,Schouteden-Nanotech-10,Aladyshkin-JPCM-20}. First, we adjust the initial height of the tip above the sample surface by varying the set-point value $I$ at a nonzero $U$. We would like to note that the absolute value of the initial height is unknown. Next, we ramp the tunneling voltage $U$ for fixed lateral coordinates of the tip, keeping the tunneling current equal to $I$. Typical acquisition time for a single spectroscopic measurement is 40 sec. During ramping, we record output signal of the active STM feedback loop $U^{\,}_{f.b.}$ (in Volts). The acquired data $U^{\,}_{f.b.}(U)$ can be transformed in so-called distance-voltage ($Z-U$) spectra: dependences of the relative height $Z=k\cdot U^{\,}_{f.b.}$ in nanometers on $U$, where $k = 26.7$\,nm/V is the conversion factor for our piezo-scanner at 78\,K. We smooth the dependence $Z(U)$ using a Gaussian filter with a window of 20\,mV to remove high-frequency noise and then numerically calculate the rate of the height variation $dZ/dU$ as a function of $U$ or given $I$.

\begin{figure*}[t!]
\centering{\includegraphics[width=8.0cm]{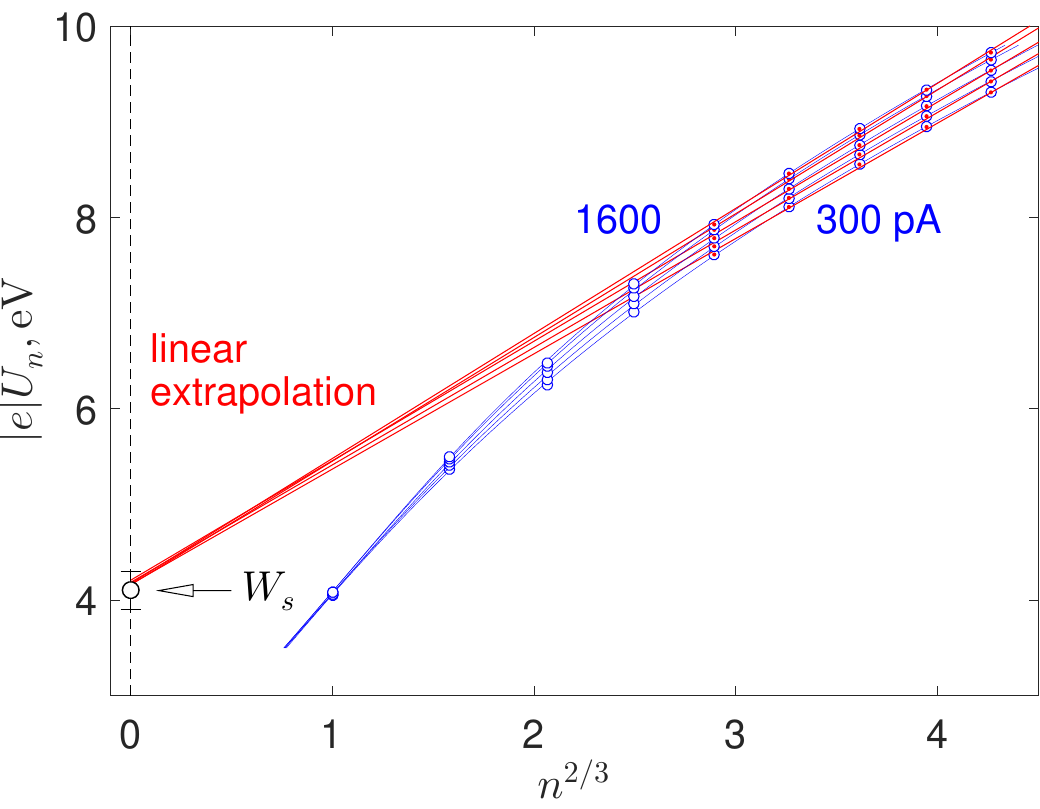}}
\caption{(color online)  Estimate for the local work function $W^{\,}_s$ for electrons of the sample based on the linear extrapolation for the dependence of the $n-$th resonant FER value on $n^{2/3}$ in the high $n-$limit considering the maxima on the dependence of $dh/dU$ on $U$ for different current: 300, 500, 800, 1200 and 1600\,pA (from bottom to top). The obtained value $W^{\,}_s\simeq 4.1 \pm 0.2$\,eV is independent on the tunneling current.}
\label{Fig06}
\end{figure*}

\begin{figure*}[t!]
\centering{\includegraphics[width=14.5cm]{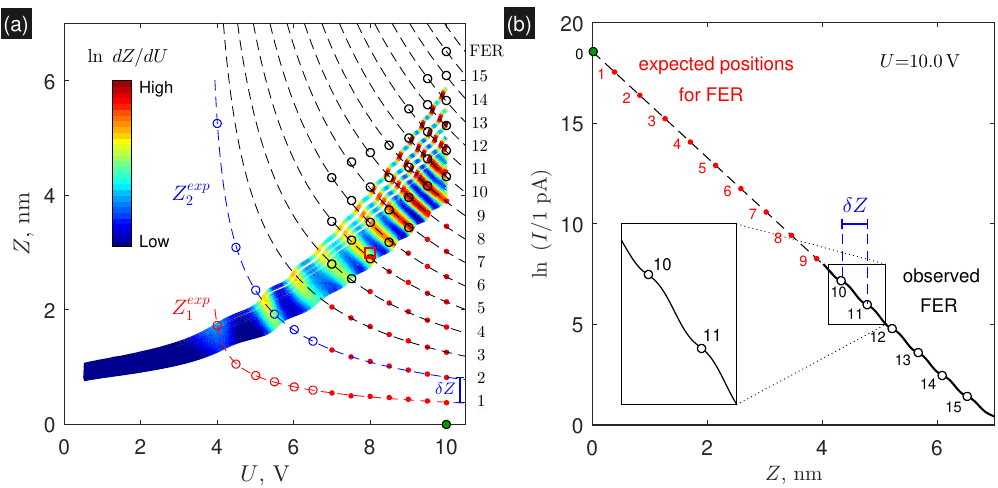}}
\centering{\includegraphics[width=12cm]{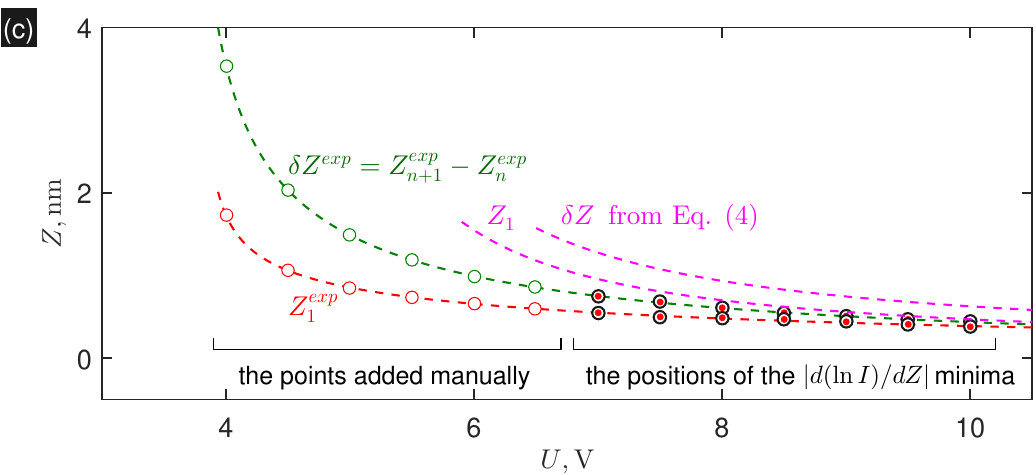}}
\caption{(color online) {\bf (a)} Series of the $Z-U$ dependences acquired for different tunneling currents: $\langle I\rangle=20$, 50, 100, 200, 300, 400, 500, 600, 800, 1000, 1200, 1400, 1600, 1800, 2200, 2600 and 3000 pA (from top to bottom). The color of the points is proportional to $\ln dZ/dU$. All these curves are slightly shifted upward and downward (about~0.05\,nm or less to compensate the vertical thermal drift and the creep effect of the piezoscanner) in such a way to get the correct dependence of $\ln I$ on $\Delta Z$ for the given bias $U^{*}$ (figure~S4 in Supporting Information); this particular adjustment is performed for $U^{*}=8$\,V, $I^{*}=1600$\,pA, $Z^{*}=3$\,nm (marked by $\Box$). The dashed lines show the series of universal hyperbolic-like dependences for the quantized values $Z^{exp}_n(U)=Z^{exp}_1(U) + (n-1)\cdot \delta Z^{exp}(U)$, corresponding to the $n-$th FER, on the mean bias $U$. \\
{\bf (b)}  The dependence of $\ln I$ on $Z$ (solid black line) at $U=10\,$V, the dashed line is the linear extrapolation for this dependence towards higher $I$ values. Black circles in the panels (a) and (b) indicate the $Z^{\,}_n$ values corresponding to the local minima in the dependences $|d(\ln\,I)/dU|$ on $Z$, where $n$ is the number of the FER, $\delta Z$ is the bias-dependent period of the oscillations of $d(\ln\,I)/dU$. Red dots in the panels (a) and (b) are the estimated positions of the local minima of $|d(\ln I)/dZ|$. \\
{\bf (c)} The dependences of $Z^{exp}_1$ on $U$ (red dashed line) and $\delta Z^{exp}=Z^{exp}_{n+1}-Z^{exp}_n$ on $U$ (green dashed line), obtained by fitting of the experimental data shown in the panel a. For the WKB model we use Eq.~(\ref{Eq:IPS-triangular-well-6}) taking into account $\Delta W=0$ and $W^{\,}_s=4.1\,$eV.
\label{Fig04}}
\end{figure*}

\section{Results and discussion}

Figure~\ref{Fig02}a shows the topography image of the Pb(111) island with two flat terraces on the upper surface. This image indicates the absence of visible and hidden defects in the vicinity of the point $\boxtimes$, where the local tunneling spectra were recorded.


The typical dependence of the rate of the height variations $dZ/dU$ on $U$ is shown in figure~\ref{Fig02}b. Periodic small-scale oscillations on the dependence $Z'(U)$ at low bias voltage ($U\lesssim 3\,$V) can be considered as clear experimental evidence for the coherent resonant tunneling through quantum-well states  (QWSs) in the thin metallic film \cite{Ustavshchikov-JETPLett-17,Altfeder-PRL-97,Su-PRL-01,Eom-PRL-06,Hong-PRB-09} (see schematics in figure~\ref{Fig01}a). Taking the period of the quantum-size oscillations $\Delta E\equiv E^{\,}_{n+1}-E^{\,}_n\simeq 0.34\,$eV, one can estimate the local thickness $D\simeq 10.9\,$nm with respect to the Si(111)$7\times 7$ surface using the relationship $D\simeq \pi\hbar v^{\,}_F/\Delta E$, where $v^{\,}_F\simeq 1.8\cdot10^{8}$\,cm/s is the Fermi velocity for the Pb(111) films \cite{Ustavshchikov-JETPLett-17,Altfeder-PRL-97}.

Aperiodic large-scale oscillations on the dependence $dZ/dU$ on $U$ clearly visible in figure~\ref{Fig02}b at high bias voltage ($U\gtrsim 3.5\,$V) point to FERs, conditioned by the coherent resonant tunneling through the surface electronic states (see schematics in figure~\ref{Fig01}b). Figure~\ref{Fig06} shows the voltage positions $U^{\,}_n$ of the FER as a function of $n^{2/3}$ extracted from the dependence of $dZ/dU$ on $U$ for different measuring current $I$, where $n$ is serial number of FER. One can see that the increase in $I$ systematically shifts all the FERs to higher $U$ values. The evolution of the resonant energies at varying $n$, $Z$ and local electric field $\mathcal{E}$ can be interpreted in terms of the quasi-classical model based on the Wentzel-Kramers-Brillouin (WKB) approximation for the electron localized in the triangular potential well (see Supporting Information). This model makes it possible to estimate the resonant bias values $U^{\,}_n$ corresponding to the higher-order FER states \cite{Aladyshkin-JPCM-20}
    \begin{eqnarray}
    \label{Eq:IPS-triangular-well-1}
    |e|U^{\,}_n \simeq W^{\,}_s + \alpha^{2/3}\cdot F^{*\,2/3}_n\cdot \left(n-\frac{1}{4}\right)^{2/3},
    \end{eqnarray}
where $n=1,\,2, ...$, $\alpha = 3\pi\hbar/(2\sqrt{2m_0}) \simeq 0.92\,{\rm nm}\cdot({\rm eV})^{1/2}$ is a constant, $F^*_n=(|e|U^{\,}_n + \Delta W)/Z$ is the gradient of the potential energy for electron above the sample surface, $\Delta W=W^{\,}_t-W^{\,}_s$ is the Volta contact potential, $\mathcal{E}^{\,}_n=-F^*_n/|e|$ is the local electrical field, $Z$ is the width of the trapezoidal potential barrier (\emph{i.e.} the distance between the STM tip and the sample surface). It is important to note, that this expression (\ref{Eq:IPS-triangular-well-1}) is identical to that derived for the positions of the $dI/dU$ maxima for the tunneling junction with trapezoidal potential barrier between two metals in the free-electron-gas approximation \cite{Kolesnychenko-PhysB-00}.

\medskip

Considering expression (\ref{Eq:IPS-triangular-well-1}) and keeping in mind that $(n-1/4)^{2/3} \simeq n^{2/3}$ for $n\gg 1$, we conclude that

\noindent (i) the difference $|e|U^{\,}_n-W^{\,}_s$ is proportional to the product of $F^{*\,2/3}_n$ and $n^{2/3}$. Since the increase in the measurement current $I$ results to the decrease in the starting height and, correspondingly, the increase in the local electric field $\mathcal{E}^{\,}_n\sim F^{*}_n \sim 1/Z$, it should cause the shift of $U^{\,}_n$ to the higher values. This conclusion is in the agreement with our findings (figure~\ref{Fig06}).

\medskip

\noindent (ii) the extrapolation of the linear fitting function for the dependence $U^{\,}_n$ on $n^{2/3}$ for the higher-order FER states towards the point $n=0$ should give a reasonable estimate for the work function $W^{\,}_s$ regardless of $W^{\,}_t$. This approach was already used for estimating the work functions for Ag and Co nanoislands \cite{Lin-PRL-07}, Pb films \cite{Aladyshkin-JPCM-20} and Pt wires \cite{Kolesnychenko-PhysB-00}.  Analysing the data in figure~\ref{Fig06}, one can see that the linear asymptotes of the dependences $U^{\,}_n$ on $n^{2/3}$ for the high-$n$ limits for different $I$ intersect the axis of ordinates at the same point. It gives us the reasonable estimate for the local work function $W^{\,}_s\simeq 4.1 \pm 0.2\,$eV for this particular Pb film independently on the measurement current.

\vspace*{3mm}

In order to study the geometrical localization of the FER we compose a 3-dimensional $Z-U-dZ/dU$ diagram that is based on individually recorded $Z-U$ spectra at various tunneling currents $I$ (figure~\ref{Fig04}a). To minimize any effect of thermal drift in the vertical direction and the vertical creep of the piezo-scanner, caused by repeating procedures of approaching/retraction of the tip, we propose the following original method of the mutual adjustment of the series of the dependences of $Z$ on $U$ acquired for the given current $I$ and the series of the dependences of $I$ on $\Delta Z$ acquired for the given voltage. We would like to emphasize that the distance $\Delta Z$ is measured from the actual position of the tip above the sample, so the absolute height of the tip above the sample surface remains unknown. Since the typical width of the here investigated atomically flat Pb(111) terraces ($\gtrsim 20\,$nm in the area of interest, see Fig.~\ref{Fig01}) is much larger than the lateral displacement of the tip during local spectral measurements ($\lesssim 1\,$nm), the lateral thermal drift can be considered as non significant for our experiments. Experimental confirmation of the independence of the FER spectrum on the tip position for the Pb(111) films (except the edges of the monatomic steps) is presented in \cite{Aladyshkin-JPCM-20}.

First, we choose the parameters $U^*$, $I^*$, $Z^*$ and assign the $Z^*$ value to the actual height $Z$ measured at $U=U^*$ and $I=I^*$. For example, the $Z-U$ diagram in figure~\ref{Fig04}a is composed for arbitrary-chosen values $U^*=8\,$V, $I^*=1600\,$pA and $Z^*=3\,$nm. Second, we record the dependence $Z(U)$ at $I=I^*$ and plot this line in the $Z-U$ diagram in such a way that this calibration curve runs through the chosen point $(U^{*}, Z^*$). Third, we record the dependence $I-\Delta Z$ at $U=U^*$. Thus, these two reference curves form a frame for the consistent visualization for the rest of the measured data in a four-dimensional space of parameters $Z-U-I-dZ/dU$ (figures\,S3 and S4 in Supporting Information). Fourth, we acquire the series of the dependences $Z$ on $U$ at $I\neq I^*$ and plot them in the pseudo-color manner choosing the color of each point proportional to $\ln\,dZ/dU$.  We check that any change in the parameters $U^*$, $I^*$ and $Z^*$ used for the adjustment does not result in a modification of the composed $Z-U$ diagram except the distortion-free shift of this diagram in the vertical direction depending on the particular $Z^*$ value.

\begin{figure*}[t!]
\centering{\includegraphics[width=8.5cm]{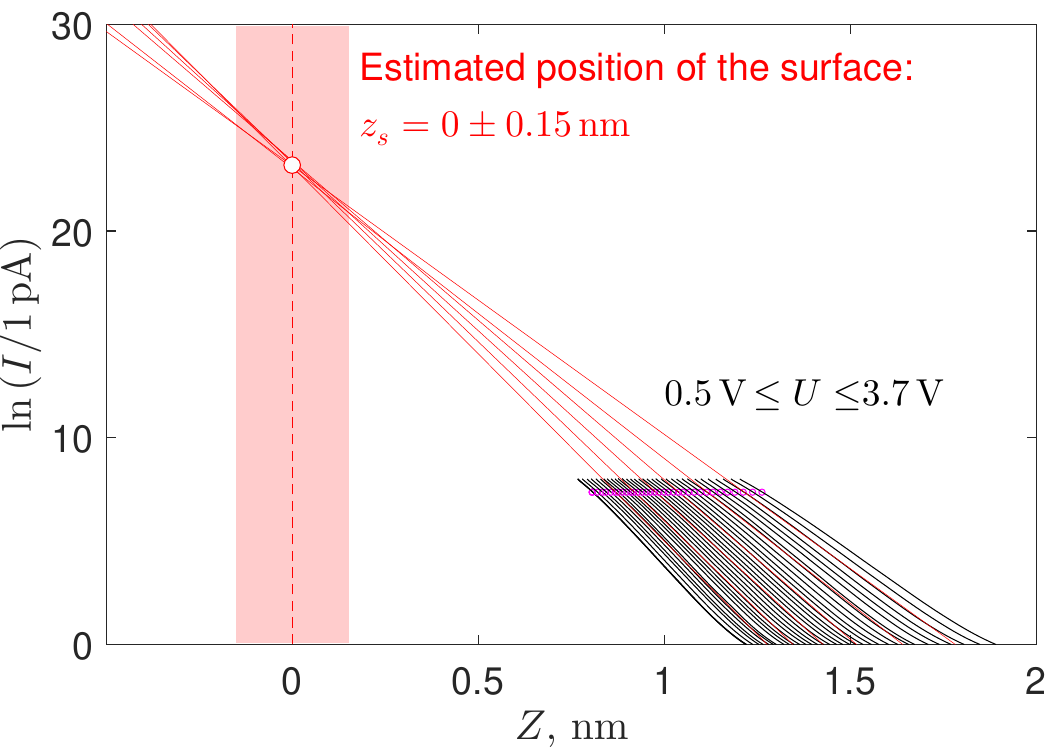}}
\caption{(color online) Typical dependences of $I$ on $Z$ in semi-logarithmic scale
and the linear extrapolation for $U = 1.0, 1.5, 2.0, 2.5, 3.0$ and 3.5\,V. The intersection point (white dot)
gives us the estimate for the position of the sample surface.
\label{Fig07-slope}}
\end{figure*}

The most remarkable feature of the $Z-U$ diagram presented in figure~\ref{Fig04}a is the well-defined stripe-like structure. Apparently, for given bias voltage ($>4\,$V) one can determine the set of the widths of the trapezoidal potential barrier $Z^{\,}_n$ in such a way that the energy of the $n-$th resonant electronic state in the triangular well will be close to the Fermi level of the tip (see figures~S5 and S6 in Supporting Information). Under this condition the electrons from the tip start to tunnel to the sample by a resonant manner, what should lead to the increase in the tunneling current for the considered quantum system. However in our case the active feedback loop immediately retract the tip from the surface in order to keep the current, what results in the rapid variations of the height at $Z\simeq Z^{\,}_n$ upon sweeping $U$. It explains why blue and red stripes in the $Z-U$ diagram do relate to the FER states. The number of the visible stripes on the $Z-U$ diagram depends on the shape of the tip (not shown in this paper), since the latter affects the profile of the triangular potential well. Interestingly that the absolute value of the slope of the red stripes in figure~\ref{Fig04}a (\emph{i. e.} $|dZ^{\,}_n/dU|$ for constant $n$) monotonously decreases as $n$ increases.

We would like to emphasize that the dependence of $\ln I$ on $Z$ could deviate from a linear decreasing function. Indeed, at high bias voltage one can see the pronounced periodic oscillations of the local slope $d(\ln I)/dZ$ as a function of $Z$. The positions of the minimal $|d(\ln\,I)/dZ|$ values are marked by black circles in figure~\ref{Fig04}b. By extrapolating the dependence of $\ln I$ on $Z$ to higher $I$ values and using the estimated period of the oscillations for the high$-n$ resonances, one can determine the expected positions of the oscillations conditioned by the low$-n$ resonances (red dots in figure~\ref{Fig04}b). We carry out this procedure for the dependences of $\ln I$ on $Z$ in the range 7\,V$\,\le U \le 10\,$V, where the periodic oscillations of the local slope are clearly visible, and thus the estimates of their periods are reliable. We mark the positions of the minimal $|d(\ln I)/dZ|$ values (both observed and expected) on the $Z-U$ diagram (figure~\ref{Fig04}a). In order to explain the periodic variations of the local slope, we rewrite Eq.~(\ref{Eq:IPS-triangular-well-1}) in the following form
    \begin{eqnarray}
    \label{Eq:IPS-triangular-well-6}
    Z^{\,}_n  \simeq \alpha\cdot \left(n-\frac{1}{4}\right)\cdot \frac{(|e|U+\Delta W)}{\left(|e|U - W^{\,}_s\right)^{3/2}},
    \end{eqnarray}
thus expressing the quantized heights $Z^{\,}_n$ as a function of $U$ and $n$. It is easy to see that in the considered model all dependences $Z^{\,}_n$ on $U$ have the universal vertical asymptote at $|e|U = W^{\,}_s$. According to Eq. (\ref{Eq:IPS-triangular-well-6}), the  difference $\delta Z^{\,}_n=Z^{\,}_{n+1}-Z^{\,}_{n}$ between two adjacent $Z^{\,}_n$ values depends only on $U$ ($\delta Z^{\,}_n\simeq \alpha/\sqrt{|e|U}$ for $|e|U \gg W^{\,}_s$) and independent of $n$ (see purple dashed line in figure figure~\ref{Fig04}c and figure\,S6 in Supporting Information). The independence of $\delta Z^{\,}_n$ on $n$ is in agreement with our experimental findings (figure~\ref{Fig04}a). Indeed, using two bias-dependent functions $Z^{exp}_1(U)$ and $\delta Z^{exp}(U)$ (red and green dashed lines in figure~\ref{Fig04}c), we can correctly describe the positions of all FERs on the $Z-U$ diagram using simple relationship: $Z^{exp}_n(U)=Z^{exp}_1(U) + (n-1)\cdot \delta Z^{exp}(U)$, where $n=1, 2, \ldots$ We conclude our simple WKB-based model correctly describes the evolution of the experimental dependences  $Z^{exp}_1(U)$ and $\delta Z^{exp}(U)$ from qualitative and quantitative (at high $U$ values) points of view.

\begin{figure*}[t!]
\centering{\includegraphics[width=8.5cm]{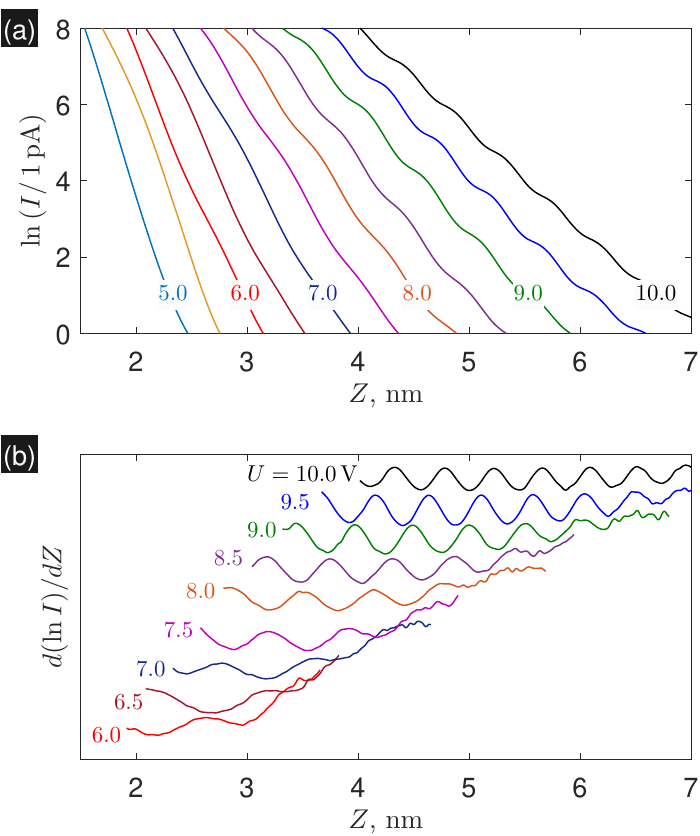}}
\caption{(color online) {\bf (a)} Adjusted dependences of $\ln I$ on $Z$ for mean bias voltage $U$ in the range from 5 to 10 V (from left to right with 0.5\,V interval). Note that the black curve for 10\,V is the same as shown in figure \ref{Fig04}b. \\
{\bf (b)} Dependences of the local slope $d(\ln\,I)/dZ$ on $Z$ for $U$ in the range from 6 to 10 V (from bottom to top); all the curves are shifted vertically for clarity.}
\label{Fig07}
\end{figure*}

The observed oscillatory dependence of $d(\ln I)/dZ$ on $Z$ makes it possible to estimate the absolute position of the STM-tip above the sample surface in a non-destructive way. Indeed, by subsequent measurements of both the $Z-U$ dependence (at constant current) and the $I-Z$ dependence (at constant high tunneling voltage) one can determine the quantized heights $Z_n$ by analyzing the variation of the slope in the $\ln I-Z$ dependence (figure~\ref{Fig04}b) and unambiguously assign the proper indices $n$ to all FER states.
The extrapolated position of the zero-order FER (green dot in figures~\ref{Fig04}a,b) yields the estimate of the z-coordinate of the sample surface. From our analysis we choose the parameter $Z^*$ to be equal to 3\,nm, since this value corresponds to the surface positioned at $z^{\,}_s\simeq 0$. An alternative way to estimate the absolute position of the sample surface is presented in figure~\ref{Fig07-slope}. This method is based on a linear extrapolation of the dependences of $\ln I$ on $Z$ for tunneling voltages below the work function to avoid the mentioned oscillations of the local slope. We find that all extrapolated functions intersect at the same point. This point can be considered as an indication for the position of the surface where a virtual electric contact takes place. Interestingly, both approaches yield the same value: $z^{\,}_s\simeq 0$. We note that, in our opinion, the accuracy of both approaches cannot be better than the typical size of the Pb atom, which is of the order of the covalent radius (0.15\,nm).

Figure~\ref{Fig07} illustrates the effect of the FERs on the current-distance dependences, which we can now understand from the analysis discussed above: (i) the appearance of the periodic oscillations on the dependence of $d\,(\ln I)/dZ$ on $Z$ at high tunneling voltage $U$ and (ii) the decrease in the period of the oscillations at increasing $U$. It should be emphasized that the dependences of $\ln I$ on $Z$ (figure~\ref{Fig07}a), used for numerical calculation of $d\,(\ln I)/dZ$ as a function of $Z$ (figure~\ref{Fig07}b), are acquired directly by means of current-distance spectroscopy without the complicated procedure of mutual alignment of the spectroscopic data. It demonstrates to that the oscillatory behaviour of $d\,(\ln I)/dZ$ as a function of $Z$ is a real physical effect inherent for all conducting samples in the field-emission regime and it cannot be considered as an artifact of data analysis. The effect of the shape of the STM tip on the period of these oscillations will be the scope of our further investigation.

\section{Conclusion}

We experimentally studied the field-emission resonances for electrons above Pb(111) films by means of low-temperature scanning tunneling spectroscopy for a broad range of tunneling conditions. The field-emission resonances can be recognized as local maxima in the dependence of the rate of the height variations $dZ/dU$ on the tunneling voltage $U$ for given $I$. Repeating single-point spectroscopical measurements for $I$ values, we composed hyperbolic-like lines on the $Z-U$ diagram illustrating the dependence of the quantized heights $Z^{\,}_n$ on $U$, where $n$ is the serial number of the field-emission resonance. We argued that the field-emission resonances strongly affect the dependence of $I$ on $Z$ for given $U$ (so-called current-distance dependence). In particular, the field-emission resonances results in nonlinear oscillatory decay of $\ln I$ and, correspondingly, non-exponential decrease of $I$ as a function of $Z$ for constant bias values exceeding the sample work function.

\section{Acknowlednements}

The authors are grateful to S. I. Bozhko and A. S. Mel'nikov for valuable comments. The work was performed with the use of the facilities at the Common Research Center 'Physics and Technology of Micro- and Nanostructures' at Institute for Physics of Microstructures RAS and funded by the Russian State Contract (No. 0030-2021-0020).

\section{Supporting Information Available}
Additional figures illustrating the principle of mutual adjustment of the tunneling spectra as well as spatial structures of the wave functions for surface electronic states are presented in the Supporting Information.

\newpage

\setcounter{page}{1}

\renewcommand{\thefigure}{S\arabic{figure}}
\renewcommand{\thepage}{S\arabic{page}}
\renewcommand{\theequation}{S\arabic{equation}}

\section*{\textcolor[rgb]{0.00,0.07,1.00}{SUPPORTING INFORMATION}}

\section*{Quantum-confined states in a one-dimensional potential well}

It is illustrative to consider a model problem describing the formation of localized electronic states in one-dimensional (1D) potential well
\begin{eqnarray}
\label{Eq01}
-\frac{\hbar^2}{2m^*}\,\frac{d^2}{dz^2}\,\psi^{\,}_n(z) + U(z)\,\psi^{\,}_n(z) = E^{\,}_n\,\psi^{\,}_n(z),
\end{eqnarray}
where $\psi^{\,}_n(z)$ and $E^{\,}_n$ are the eigenfunctions and the eigenvalues of the Schr\"{o}dinger equation (\ref{Eq01}) and $U(z)$ is the potential energy.
For simplicity we take the periodic potential inside the quantum well in the form of a combination of two Fourier components
    \begin{eqnarray}
    \label{Eq02}
    U(z)=\left\{%
    \begin{array}{lc}
    U^{\,}_{left} & \hbox{\quad for $z < - L/2$;} \\
    U^{\,}_0 + U^{\,}_1\cdot\cos\left(2\pi z/a\right) + U^{\,}_2\cdot\cos\left(4\pi z/a\right) & \hbox{\quad for $-L/2\le z\le L/2$;} \\
    U^{\,}_{right} & \hbox{\quad for $z>L/2$,} \\
    \end{array}%
    \right.
    \end{eqnarray}
where $a$ is the lattice parameter and $L$ is the width of the quantum well. Two Fourier-components of the potential ensure the formation of two forbidden gaps in the spectrum of the electronic eigenstates.
\begin{figure*}[h!]
\centering{\includegraphics[width=14cm]{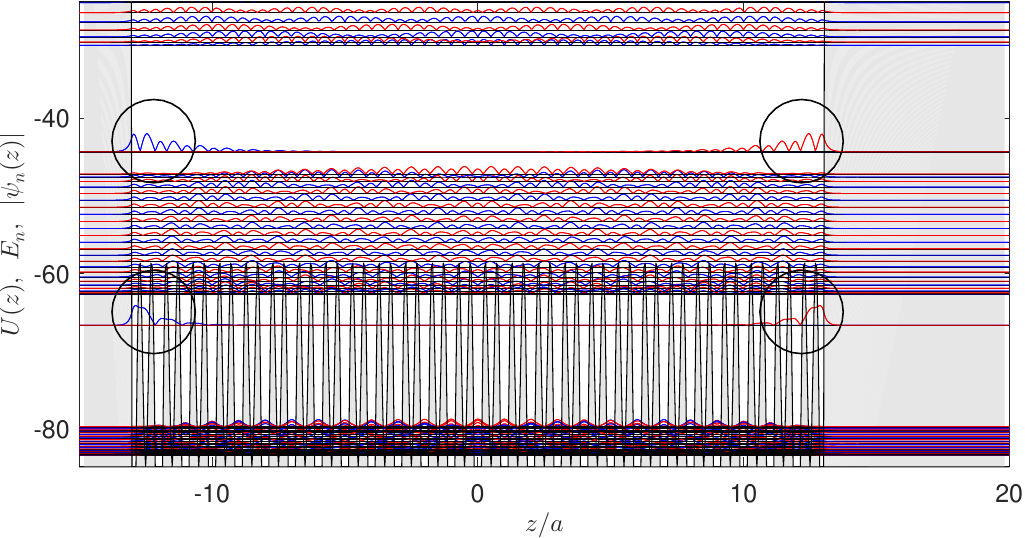}}
\caption{Examples of the stationary eigenfunctions $|\psi^{\,}_n(z)|$ of Eq.~(\ref{Eq01}) for a one-dimensional quantum well with almost symmetric edges: $U^{\,}_0=-8\,E^{\,}_0$, $U^{\,}_1=-0.5\,E^{\,}_0$ and $U^{\,}_2=-0.5\,E^{\,}_0$, where $E^{\,}_0=\pi^2\hbar^2/(2m^{\,}_0a^2)$ is the natural energy scale. All these eigenfunctions are shifted vertically for clarity in accordance to their eigenvalues $E^{\,}_n$. A concentration of the eigenvalues points to the allowed energy bands. Large circles mark the Tamm-Shockley surface states lying within the first and the second forbidden gaps of the bulk crystal.}
\label{Fig00-sm}
\end{figure*}

\begin{figure*}[h!]
\centering{\includegraphics[width=14cm]{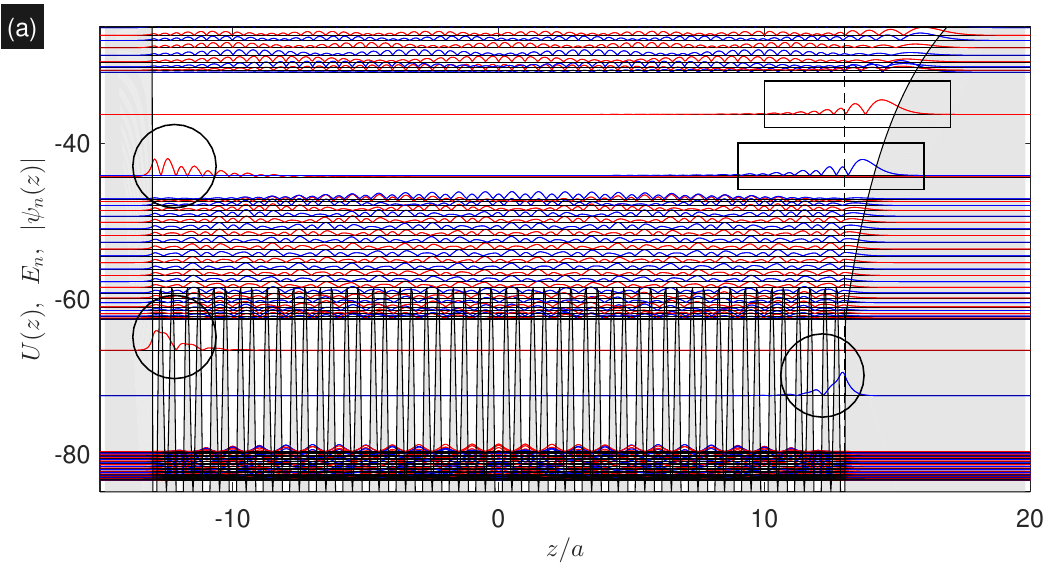}}
\centering{\includegraphics[width=14cm]{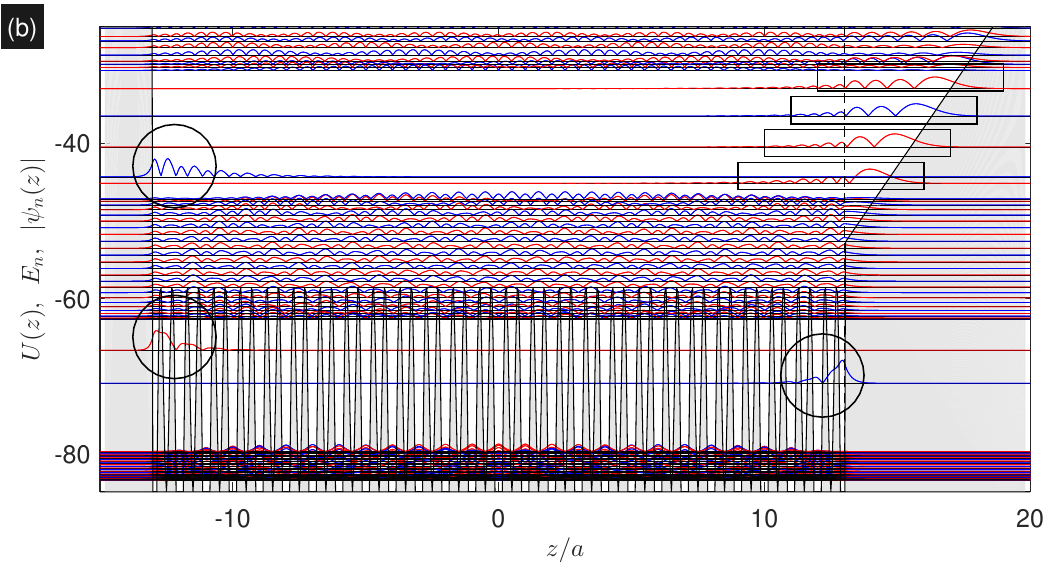}}
\caption{{\bf (a, b)} Examples of the stationary eigenfunctions $|\psi^{\,}_n(z)|$ of Eq.~(\ref{Eq01}) for a one-dimensional quantum well with non-symmetric edges and Coulomb potential near the right edge (panel a) or linearly increasing electric potential near the right edge (panel b): $U^{\,}_0=-8\,E^{\,}_0$, $U^{\,}_1=-0.5\,E^{\,}_0$ and $U^{\,}_2=-0.5\,E^{\,}_0$, where $E^{\,}_0=\pi^2\hbar^2/(2m^{\,}_0a^2)$. All these eigenfunctions are shifted vertically for clarity. Large circles mark the classical Tamm-Shockley surface states lying within the first and the second forbidden gaps of the bulk crystal, while rectangles depict the image-potential states (panel a) and the surface states localized in a linearly increasing potential (panel b) in the second forbidden gap.}
\label{Fig00ab-sm}
\end{figure*}

The results of the direct numerical solution of Eq.~(\ref{Eq01}) for the potential well with almost symmetrical edges ($U^{\,}_{left}=0$ and $U^{\,}_{right}\simeq 0$) are presented in figure~\ref{Fig00-sm}. One can see that the periodic potential induces well-known Tamm-Shockley surface states localized near the edges of the crystal with energies lying in the forbidden gaps of the bulk crystal. Figure \ref{Fig00ab-sm} demonstrates the modification of both eigenfunctions and eigenvalues caused by a Coulomb-like potential near the right edge of the crystal ($U^{\,}_{right} \simeq {\rm const}^{\,}_1 - {\rm const}^{\,}_2/(z-L/2)$, panel a) and by a linearly increasing potential near the right edge of the crystal ($U^{\,}_{right} \simeq {\rm const}^{\,}_1 + {\rm const}^{\,}_2\cdot (z-L/2)$, panel b).

These examples illustrate that all types of the considered surface states (namely, classical Tamm-Shockley states, image-potential states and states localized in a linearly increasing potential) are equivalent from the physical point of view. In addition, it is clear that the energies and the spatial structure of the wave functions of the localized states are very sensitive to the profile of the potential energy outside the potential well.

\newpage

\section*{Mutual adjustment of the $Z-U$ and $I-\Delta Z$ dependences}

\begin{figure*}[h!]
\centering{\includegraphics[width=17.5cm]{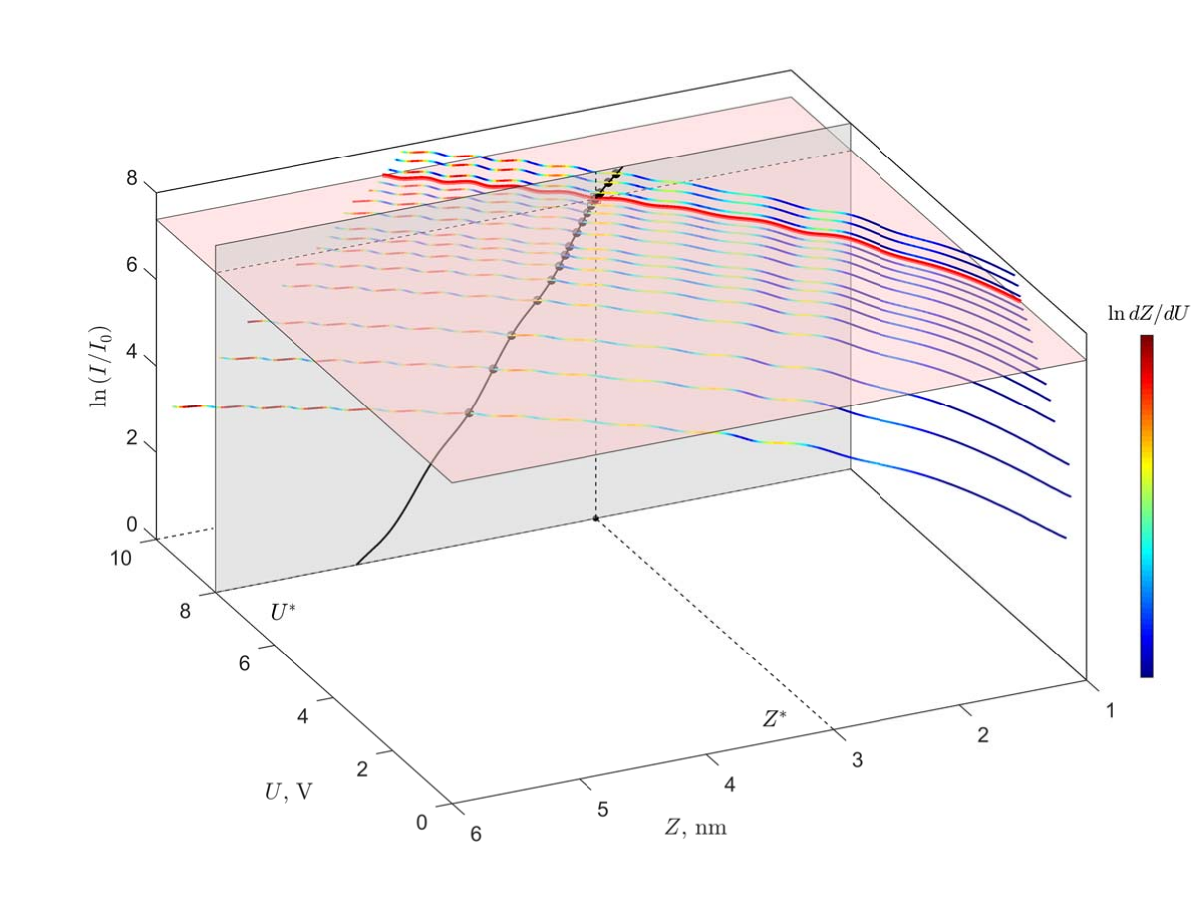}}
\caption{Diagram illustrating the principle of the mutual adjustment of the $Z-U$ dependences acquired in the regime of constant current and $I-Z$ dependences acquired in the regime of constant voltage in a 3D geometry. This particular adjustment is carried out for the following values: $U^{*}=8$\,V, $I^{*}=1600$\,pA, $Z^{*}=3$\,nm, $I^{\,}_0=1\,$pA. The color of all points is proportional to $\ln dZ/dU$. The same plot in 2D geometry is presented in Figure~\ref{Fig03sm} and Figure~4a of the main paper.}
\label{Fig01sm}
\end{figure*}

\begin{figure*}[t!]
\centering{\includegraphics[width=16cm]{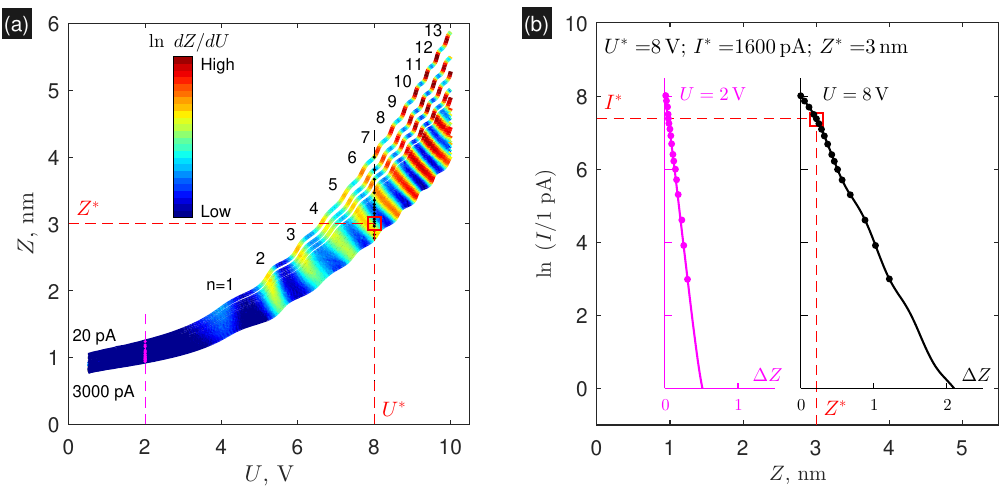}}
\caption{Diagrams illustrating the principle of the mutual adjustment of the $Z-U$ dependences acquired in the regime of constant current and $I-\Delta Z$ dependences acquired in the regime of constant voltage: \\
{\bf (a)} Series of the $Z-U$ dependences acquired for different tunneling currents: $I=20$, 50, 100, 200, 300, 400, 500, 600, 800, 1000, 1200, 1400, 1600, 1800, 2200, 2600 and 3000 pA (from top to bottom). The color of the points is proportional to $\ln dZ/dU$. All these curves are slightly shifted upward and downward (about~0.05\,nm or less to compensate vertical thermal drift and creep effect of the piezoscanner) in such a way to get the correct dependence of $\ln I$ on $\Delta Z$ for the given bias $U^{*}$ (black dots in the panel b); this particular adjustment is performed for $U^{*}=8$\,V, $I^{*}=1600$\,pA, $Z^{*}=3$\,nm (marked by $\textcolor[rgb]{1.00,0.00,0.00}{\Box}$). \\
{\bf (b)} Purple and black solid lines are the dependences of $\ln\,I$ on $\Delta Z$, acquired during current-distance spectroscopy measurements for $U=+2$ and +8\,V, correspondingly. Purple and black dots represent the cross-sectional views for the data shown in the panel a along the dashed vertical lines for $U=+2$ and +8~V, respectively. Almost perfect agreement between different kinds of measurements in panel b ensures the correctness of the visualization of the data in panel a.}
\label{Fig03sm}
\end{figure*}

\newpage

\,\,

\newpage

\section*{The estimate of the FER energies based on the quasi-classical Wentzel-Kramers-Brillouin approximation}

We consider the simplest quasiclassical 1D model and calculate the energy and spatial structure of the localized states in the linearly increasing potential. This model was partly presented in our recent paper (Aladyshkin, J. Phys.: Condens. Matt. {\bf 32}, 435001 (2020)).
This consideration is relevant for the particular cases of (i) a planar structure with a tunneling barrier of constant width and (ii) a very blunt STM tip above the sample surface. This model seems to be helpful for investigating the periodic nature of the quantized heights $Z^{\,}_n$ attributed to the $n-$th field-emission resonances.


\medskip

\begin{figure*}[b!]
\centering{\includegraphics[width=15cm]{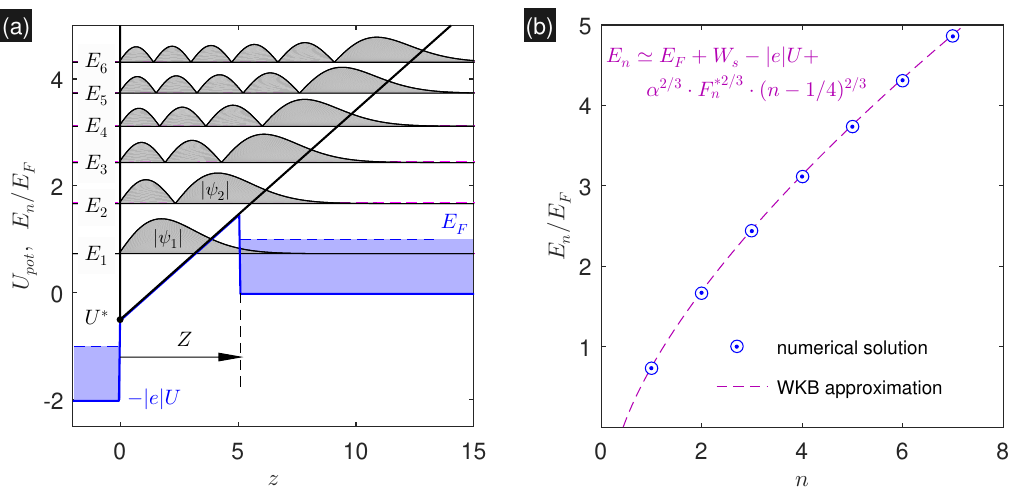}}
\caption{{\bf (a)} The model profiles of the potential energy in the form of the trapezoidal barrier $U^{(1)}_{pot}(z)$ (blue solid line) and triangular potential well $U^{(2)}_{pot}(z)$ (black solid line). Thin black and purple horizontal lines are the eigenenergies of Eqs.~(\ref{Eq:IPS-triangular-well-21a})--(\ref{Eq:IPS-triangular-well-21b}) obtained numerically and analytically, respectively, for the following parameters $W^{\,}_s/E^{\,}_F=0.5$, $|e|U/E^{\,}_F=2$ and $\Delta W=0$. The difference between the analytical and numerical solutions for $E^{\,}_n$ are less than 2\% and practically invisible in this scale.
The gray shaded oscillatory patterns are the eigenfunctions of the problem (\ref{Eq:IPS-triangular-well-21a})--(\ref{Eq:IPS-triangular-well-21b}) calculated numerically for an electron in the triangular potential well. \\
{\bf (b)} Comparison of the eigenenergies $E^{\,}_n$ obtained analytically [dashed line, Eq.~(\ref{Eq:IPS-triangular-well-10})] and numerically [$\textcolor[rgb]{0.00,0.07,1.00}{\odot}$, Eqs.~(\ref{Eq:IPS-triangular-well-21a})--(\ref{Eq:IPS-triangular-well-21b})] for $W^{\,}_s/E^{\,}_F=0.5$, $|e|U/E^{\,}_F=2$ and $\Delta W=0$.}
\label{Fig02sm-1}
\end{figure*}

The reflection coefficient for an electron from a flat surface of the sample apparently depends on the energy. The amplitude of the reflection coefficient can be close to unity if $E$ lies within the forbidden energy gap of the bulk crystal. Such almost ideal  reflection ability in fact leads to the formation of the slow-decaying quasistationary electronic states localized above the sample in a linearly increasing potential. We therefore consider an effective triangular potential well (black solid line in Figure~\ref{Fig02sm-1}a)
    \begin{eqnarray}
    \label{PotentialEnergy-triangular}
    U^{(2)}_{pot}(z) = \left\{%
    \begin{array}{ll}
    \infty           & \hbox{for $z<0$,} \\
    U^* + F^*\cdot z & \hbox{for $z\ge 0$;} \\
    \end{array}%
    \right.
    \end{eqnarray}
instead of a trapezoidal potential barrier
    \begin{eqnarray}
    \label{PotentialEnergy-trapezia}
    U^{(1)}_{pot}(z) =\left\{%
    \begin{array}{ll}
    -|e|U            & \hbox{for $z<0$;} \\
    U^* + F^*\cdot z & \hbox{for $0\le z\le Z$;} \\
    0                & \hbox{for $z>Z$.} \\
    \end{array}%
    \right.
    \end{eqnarray}
Here $E^{\,}_F$ is the Fermi energy of both electrodes (since the sample and the tip are in electrical contact,the Fermi levels of the electrodes are at the same energy); $U$ is the electric potential of the sample with respect to the tip, $U^*=E^{\,}_F+W^{\,}_s-|e|U$ is the bias--shifted potential energy at the sample surface (at $z\to 0^+$); $W^{\,}_s$ and $W^{\,}_t$ are the work function for electrons of the sample and the tip, respectively; $F^*=(|e|U + \Delta W)/Z$ is the gradient of the potential energy which is proportional to the effective electric field above the surface; $\Delta W=W^{\,}_t-W^{\,}_s$ accounts for the Volta contact potential due to the difference in the work functions; and $Z$ is the distance between the STM tip and the sample surface.

\vspace*{0.5cm}

We start with the substitution of the quasi-classical momentum $p(z)=\sqrt{2m^{\,}_0\,(E-U^{(2)}_{pot}(z))}$ into the Bohr-Sommerfeld quantization rule
    \begin{eqnarray}
    \label{Bohr-Sommerfeld}
    \frac{1}{\pi\hbar}\,\int\limits_{a}^{b} p(z)\,dz = m + \gamma, \qquad m = 0, 1, \ldots
    \end{eqnarray}
where $a=0$ and $b=(E-U^*)/F^{*}$ are classical turning points, $m^{\,}_0$ is the mass of free electron, $m$ is the integer index, and the numerical coefficient $\gamma$ is close to 3/4 due to the presence of the impenetrable barrier at $z=0$. We introduce the index $n=m+1$ for convenience. After integration, we obtain the discrete energy spectrum for the electrons localized in the triangular potential well
    \begin{eqnarray}
    \nonumber
    E^{\,}_{n} = U^* + \left\{\frac{3}{2}\,\frac{\pi\hbar}{\sqrt{2m^{\,}_{0}}}\,F^{*}_n\cdot\left(n - \frac{1}{4}\right)\right\}^{2/3} = \\ \qquad =  E^{\,}_F + W^{\,}_s - |e|U + \left\{\frac{3}{2}\,\frac{\pi\hbar}{\sqrt{2m^{\,}_{0}}}\,F^{*}_n\cdot \left(n - \frac{1}{4}\right)\right\}^{2/3}, \qquad n = 1, 2, \ldots
    \label{Eq:IPS-triangular-well-2}
    \end{eqnarray}

\medskip
\noindent
Since the process of resonant tunnelling starts when the energy of the state localized in the area between the emitter (tip) and collector (sample) reaches the Fermi energy of the tip ($E^{\,}_{n}\simeq E^{\,}_{F}$), we come to the relationship
    \begin{eqnarray}
    \label{Eq:IPS-triangular-well-3}
    |e|U^{\,}_n \simeq W^{\,}_s + \left(\frac{3}{2}\,\frac{\pi\hbar}{\sqrt{2m^{\,}_{0}}}\right)^{2/3}\,F^{*\,2/3}_n\,\left(n-\frac{1}{4}\right)^{2/3}, \quad n=1, 2, \ldots
    \end{eqnarray}

\vspace*{0.3cm}
The quasiclassical wave function of the electron localized in the triangular potential well near the right turning point $z=b$ can be written in the following form
    \begin{eqnarray}
    \nonumber \psi^{\,}_n(z)\simeq \frac{C}{\sqrt{p(z)}}\,\cos\left(\frac{1}{\hbar}\,\int\limits_z^b p(z)\,dz - \frac{\pi}{4}\right) \simeq \\ \quad \simeq \frac{C}{\sqrt{2m^{\,}_0\,(E^{\,}_n-U^*-F^*_nz)}}\,\cos\left(\frac{2}{3F^*_n}\,\frac{\sqrt{2m^{\,}_0}}{\hbar}\,(E^{\,}_n-U^*-F^*_nz)^{3/2} - \frac{\pi}{4}\right),
    \label{Eq:IPS-triangular-well-4}
    \end{eqnarray}
where $C$ is a constant.
It is easy to see that the wave function at the left turning point $z=0$ is equal to zero provided that the argument of the cosine function is equal to $-\pi/2 + \pi\,n$
    \begin{eqnarray}
    \nonumber \frac{2}{3}\,\frac{\sqrt{2m^{\,}_0}}{\hbar F^*_n}\,(E^{\,}_n-U^*)^{3/2} - \frac{\pi}{4} \simeq -\frac{\pi}{2} + \pi\,n.
    \end{eqnarray}
This equation is equivalent to Eq.~(\ref{Eq:IPS-triangular-well-2}).

\vspace*{0.5cm}
For convenient comparison of the analytic expression (\ref{Eq:IPS-triangular-well-3}) with the results of numerical simulations we introduce the following dimensionless parameters: the reduced energy $\varepsilon=E/E^{\,}_F$, the reduced bias voltage $u=|e|U/E^{\,}_F$, the reduced work function $w^{\,}_s=W^{\,}_s/E^{\,}_F$ and $\Delta w=\Delta W^{\,}_s/E^{\,}_F$ as well as the reduced coordinate $z'=z/L$ and the reduced barrier width $h'=Z/L$, where $E^{\,}_F$ is the natural energy scale and $L=\sqrt{\hbar^2/(2m^{\,}_0E^{\,}_F)}$ is the length scale. The expressions (\ref{Eq:IPS-triangular-well-2}) and (\ref{Eq:IPS-triangular-well-3}) can be written in the dimensionless form
    \begin{eqnarray}
    \label{Eq:IPS-triangular-well-10}
    \varepsilon^{\,}_{n} = 1 + w^{\,}_s - u +  \left(\frac{3}{2}\,\pi\right)^{2/3}\,\frac{(u+\Delta w)^{2/3}}{h^{\prime\,2/3}}\,\left(n- \frac{1}{4}\right)^{2/3}.
    \end{eqnarray}
and
    \begin{eqnarray}
    \label{Eq:IPS-triangular-well-11}
    u^{\,}_n \simeq w^{\,}_s +  \left(\frac{3}{2}\,\pi\,\right)^{2/3}\,\frac{(u^{\,}_n+\Delta w)^{2/3}}{h^{\prime\,2/3}_n}\,\left(n- \frac{1}{4}\right)^{2/3}.
    \end{eqnarray}
By expressing $h'_n$ via $u^{\,}_n$ we get the dimensionless relationship for the $n-$th resonant value for the barrier width as a function of the bias voltage
    \begin{eqnarray}
    \label{Eq:IPS-triangular-well-12}
    \frac{Z^{\,}_n}{L} \simeq \frac{3}{2}\,\pi\,\frac{(u^{\,}_n+\Delta w)}{(u^{\,}_n - w^{\,}_s)^{3/2}}\,\left(n-\frac{1}{4}\right), \qquad n = 1, 2, \ldots
    \end{eqnarray}
This implies that in this model the ratios $h^{\,}_1:h^{\,}_2:h^{\,}_3 \ldots$ for a given voltage have the universal form $3 : 7 : 11 \ldots$

\vspace*{0.5cm}
Finally, the derived Schrodinger equation for determination of the real-valued eigenfunctions $\psi^{\,}_n(z)$ and eigenvalues $E^{\,}_n$ for a particle trapped in the linearly increasing potential $U^{(2)}_{pot}$ can be formulated in the dimensionless form as follows
\begin{eqnarray}
\label{Eq:IPS-triangular-well-21a}
-\frac{d^2}{dz^{\prime\,2}}\,\psi^{\,}_n(z') + \left\{(1 + w^{\,}_s - u) + (u+\Delta w)\cdot\frac{z'}{h'} \right\}\,\psi^{\,}_n(z') = \varepsilon^{\,}_n\,\psi^{\,}_n(z'),
\end{eqnarray}
\begin{eqnarray}
\label{Eq:IPS-triangular-well-21b}
\mbox{Boundary conditions:} \quad \psi^{\,}_n(z')\Big|^{\,}_{z'=0}=0 \quad \mbox{and} \quad \psi^{\,}_n(z')\Big|^{\,}_{z'=\infty}=0.
\end{eqnarray}

\begin{figure*}[h!]
    \centering{\includegraphics[width=16.5cm]{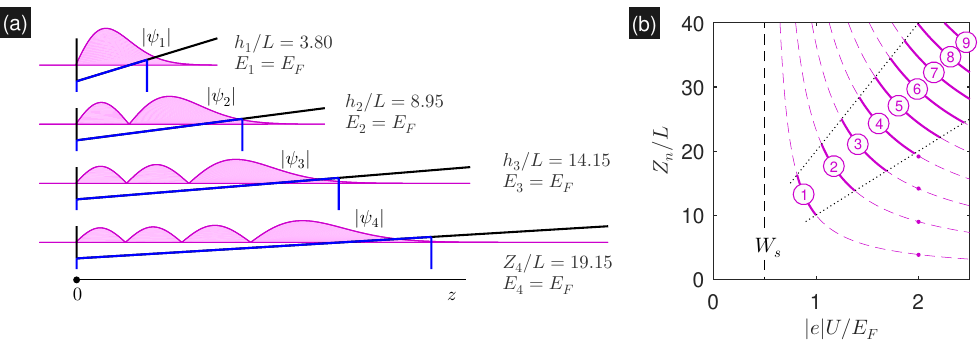}}
    \caption{ {\bf (a)}  The pink shaded oscillatory patterns are the eigenfunctions of the Schr\"{o}dinger equation Eqs.~(\ref{Eq:IPS-triangular-well-21a})--(\ref{Eq:IPS-triangular-well-21b}) satisfying the criterion of the resonant tunneling $E^{\,}_n\simeq E^{\,}_F$. \\
    {\bf (b)} The normalized quantized heights $Z^{\,}_n/L$ as a function of the dimensionless bias $|e|U/E^{\,}_F$ obtained analytically by means of Eq.~(\ref{Eq:IPS-triangular-well-12}), where we take $W^{\,}_s/E^{\,}_F=0.5$ and $\Delta W = 0$. Two dotted lines depict schematically the range of experimentally accessible current values (compare this plot to figure~4a of the main paper).}
    \label{Fig02sm-2}
    \end{figure*}

\vspace*{0.5cm}
The results of the numerical solution of Eqs.~(\ref{Eq:IPS-triangular-well-21a}) and (\ref{Eq:IPS-triangular-well-21b}) for $E^{\,}_n=E^{\,}_F$ are presented in Figures~\ref{Fig02sm-1}a and \ref{Fig02sm-2}a. The evolution of the quantized heights $Z^{\,}_n$ as a function of the voltage bias, described by Eq.~(\ref{Eq:IPS-triangular-well-12}),  is shown in Figure~\ref{Fig02sm-2}b.

\end{document}